\documentclass[a4paper,11pt]{article}
\usepackage{pos}

\usepackage{booktabs}
\usepackage{tabularx}

\title{Impact of LHC jet data on Parton Distribution Functions}

\author*[a]{Jo\~{a}o Pires}

\affiliation[a]{LIP,
  Avenida Professor Gama Pinto 2, P-1649-003, Lisbon, Portugal}

\emailAdd{jnpires@lip.pt}

\abstract{In this talk we discuss recent results on the impact of LHC jet data on global fits of parton distribution functions using theoretical predictions at NNLO in QCD supplemented by electroweak (EW) corrections.}

\FullConference{%
  The Eighth Annual Conference on Large Hadron Collider Physics-LHCP2020 \\
  25-30 May, 2020\\
  online}

\begin{document}
\maketitle

\section{Introduction}
Single-inclusive jet and dijet observables are the most fundamental QCD processes measured at hadron colliders. They probe the basic parton-parton scattering and thus allow for a determination of the parton distribution functions in the proton. The calculation of next-to-next-to-leading order (NNLO) QCD corrections to jet cross sections at hadron colliders was completed recently~\cite{Ridder:2013mf,Currie:2016bfm,Currie:2017eqf,Gehrmann-DeRidder:2019ibf} (see also Ref.~\cite{Czakon:2019tmo}), and opened up the possibility of doing precision phenomenology with jet observables. 

In this talk we discuss recent results~\cite{AbdulKhalek:2020jut} on the impact of adding LHC jet data to a global fit of parton distribution functions within the framework of the NNPDF3.1 global PDF determination~\cite{Ball:2017nwa}, using the most up-to-date jet predictions through NNLO QCD calculated with NNLOJET~\cite{Currie:2018oxh} combined with NLO EW corrections~\cite{Dittmaier:2012kx}. We will consider separately the impact of the complete single-inclusive jet~\cite{Aad:2014vwa,Aaboud:2017dvo,
Chatrchyan:2012bja,Khachatryan:2016mlc}
and dijet~\cite{Aad:2013tea,Chatrchyan:2012bja,Sirunyan:2017skj} datasets
from ATLAS and CMS at $\sqrt{s}=7$ and 8 TeV. For both jet observables we will look at the $\chi^2$ of the theoretical prediction for each dataset or combinations of datasets, defined according to  Eqs.~(7)-(8) of Ref.~\cite{Ball:2012wy}.

\section{Impact of single-inclusive jet data}
\begin{table}[b]
\renewcommand*{\arraystretch}{1.60}
\scriptsize
\centering
\begin{tabularx}{\textwidth}{Xrccccccccc}
\toprule
 Dataset                    & $n_{\rm dat}$ &     b  &   bn   &  janw  &    j7  &   j7n  &  j7nw  &    j8  &   j8n  &  j8nw  \\
\midrule
 Jets (all)                 &        520  & [1.48] & [2.60] &  1.88  & [1.86] & [2.45] & [2.53] & [1.20] & [1.75] & [1.89] \\
 \ \ Jets (fitted)          &             &  ---   &  ---   &  1.88  &  0.79  &  1.15  &  1.12  &  1.40  &  2.05  &  2.20  \\
 \ \ ATLAS 7 TeV            &         31  & [1.26] & [1.87] &  1.59  &  1.12  &  1.73  &  1.15  & [1.07] & [1.69] & [1.62] \\
 \ \ ATLAS 8 TeV            &        171  & [2.60] & [5.01] &  3.22  & [3.55] & [4.76] & [4.58] &  2.03  &  3.18  &  3.25  \\
 \ \ CMS   7 TeV            &        133  & [0.60] & [1.06] &  1.09  &  0.71  &  1.01  &  1.11  & [0.72] & [0.94] & [1.14] \\
 \ \ CMS   8 TeV            &        185  & [1.10] & [1.59] &  1.25  & [1.24] & [1.47] & [1.80] &  0.81  &  1.01  &  1.23  \\
 Dijets (all)               &        266  & [3.49] & [3.07] & [2.10] & [4.16] & [2.96] & [2.56] & [3.34] & [2.21] & [2.22] \\
 \ \ Dijets (fitted)        &             &  ---   &  ---   &  ---   &  ---   &  ---   &  ---   &   ---  &  ---   &  ---   \\
 \ \ ATLAS 7 TeV            &         90  & [1.49] & [2.47] & [1.95] & [1.77] & [2.46] & [1.97] & [1.43] & [2.28] & [2.01] \\
 \ \ CMS   7 TeV            &         54  & [2.06] & [2.40] & [2.08] & [2.43] & [2.50] & [2.12] & [1.65] & [2.00] & [2.15] \\
 \ \ CMS   8 TeV            &        122  & [5.60] & [3.81] & [2.21] & [6.70] & [3.53] & [3.20] & [5.48] & [2.26] & [2.39] \\
\midrule
 Total                      &             &  1.20  & 1.18   &  1.28  &  1.17  &  1.17  & 1.17  &   1.39  &  1.27  &  1.27  \\
\bottomrule
\end{tabularx}

\vspace{0.3cm}
\caption{The $\chi^2$ per datapoint for all fits including single-inclusive jet data.
  Results are presented both for the sets included in each fit, and also for those not
  included, enclosed in square brackets. For each dataset the number of datapoints is also shown~\cite{AbdulKhalek:2020jut}.}
\label{tab:chi2s}
\end{table}

We first present PDF fits obtained by adding single-inclusive jet data to the baseline fit (labelled \#bn) of the NNPDF3.1 analysis~\cite{Ball:2017nwa}, using either the full data set, or 7 TeV data or 8 TeV data only, and with theory at pure NLO QCD, pure NNLO QCD, or NNLO QCD supplemented by EW corrections. The values of the $\chi^2$ per datapoint for all fits are collected in Table~\ref{tab:chi2s}. 

From the table we can observe that in the fit of the combined ATLAS and CMS 7 and 8 TeV datasets at NNLO QCD+NLO EW (labelled \#janw), individual jet datasets are well described (with $\chi^2$ per datapoint of order one), except the 8 TeV ATLAS data ($\chi^2$ = 3.22). In~\cite{AbdulKhalek:2020jut} it is shown that an alternative choice of correlation model for this dataset significantly improves the fit quality without a significant change in PDFs. Finally, we can observe an improved description of dijet data, which are not included in any of these fits, with respect to the baseline fit \#bn. This suggests that the inclusion of single-inclusive and dijet data have a similar impact on PDFs.

\begin{figure}[!t]
\centering
\includegraphics[scale=0.45]{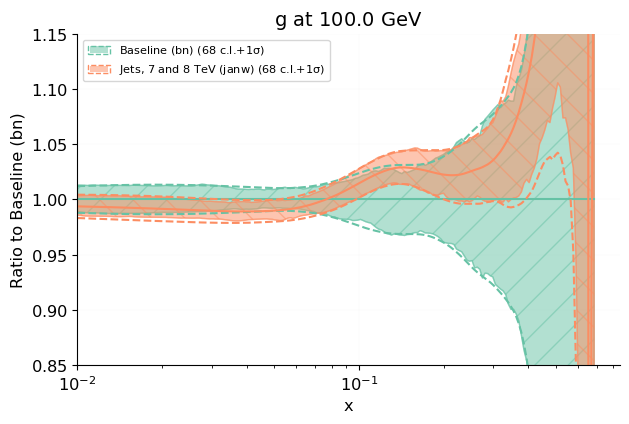}
\includegraphics[scale=0.45]{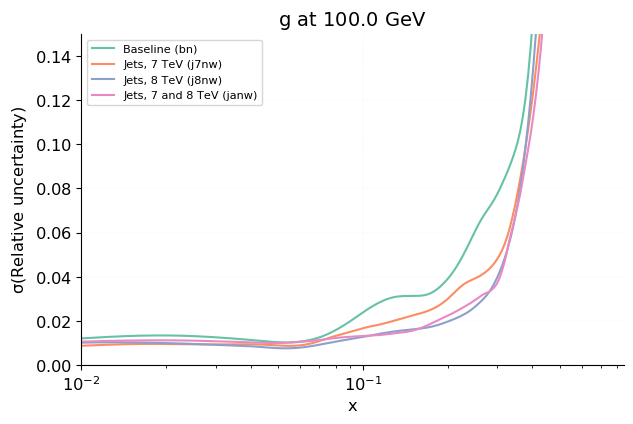}\\
\caption{Comparison between the baseline fit with no jet data (\#bn)
  and the fit with all single-inclusive jet data included (\#janw) (left).  The relative uncertainty on the gluon PDF (right) at $Q=100$ GeV~\cite{AbdulKhalek:2020jut}.}
\label{fig:jet_data_total}
\end{figure}

For the case of a global NNPDF3.1-like PDF determination we observed that the inclusion of the single-inclusive jet data essentially impacts only the gluon PDF. Quantitatively we can observe in Fig.~\ref{fig:jet_data_total}, a suppression of the gluon density of 2\% in the small-$x$ region and an enhancement of 4\% in the large x-region, within the
uncertainty of the baseline PDF. Moreover, the impact of the data can also be seen in a reduction of gluon uncertainty at x$\approx$0.2 from 4\% to 1.5\%, mostly driven by the inclusion of the 8 TeV data~\cite{AbdulKhalek:2020jut}. 

\section{Impact of dijet data}
To assess the impact of dijet data on PDFs we compare fits with optimal settings, i.e. with
NNLO QCD theory, and EW corrections included, and either the
full dataset (\#danw), or the 7~TeV (\#d7nw) or 8~TeV (\#d8nw) data included in turn.

From Table~\ref{tab:chi2d}, we observe that individual dijet datasets are generally reasonably well described (the $\chi^2$ per datapoint is around 1.5 for each). Moreover, after the inclusion of the dijet data in the baseline fit we observe an improvement in the description of the single-inclusive jet data. This result indicates consistency between the single-inclusive and dijet data. However, we note that contrary to the case of single-inclusive jet data, no tension is seen between dijet data and the rest of the global dataset (specifically top rapidity distributions), whose $\chi^2$ is left almost unaffected~\cite{AbdulKhalek:2020jut}.

\begin{figure}[!b]
\centering
\includegraphics[scale=0.46]{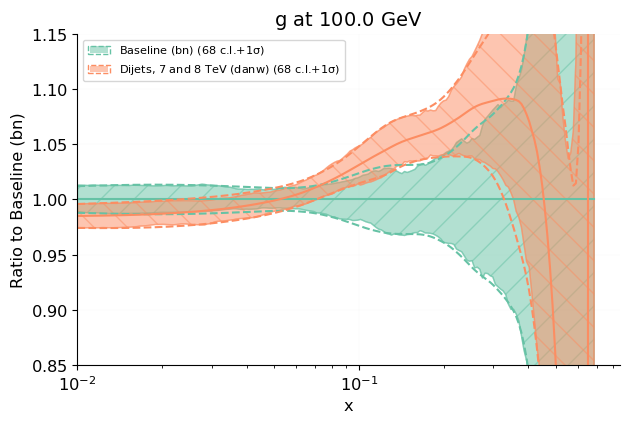}
\includegraphics[scale=0.46]{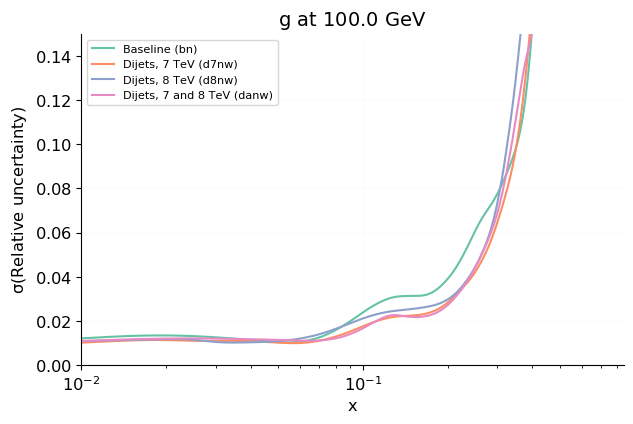}\\
\caption{Same as Fig.~\ref{fig:jet_data_total}, but now for dijets~\cite{AbdulKhalek:2020jut}.}
\label{fig:dijet_total}
\end{figure}

The relative impact of the dijet data on the gluon central values and uncertainty
can be directly inferred from Fig.~\ref{fig:dijet_total}. With respect to the baseline fit we observe a suppression of the gluon density by 2\% in the small-$x$ region and an enhancement by 10\% at $x\approx0.3$ upon inclusion of dijet data. Therefore, we conclude that we observe qualitatively similar pulls on the gluon distribution as we did with single inclusive jet data. However, the availability of high precision triple differential measurements of the dijet cross section at 8 TeV by the CMS collaboration~\cite{Sirunyan:2017skj}, leads to stronger constraints on the gluon distribution and in a wider kinematical region.

The reduction in uncertainty in the gluon distribution in comparison to the baseline (Fig.~\ref{fig:dijet_total}), upon inclusion of the dijet data, is about 3-4\% reaching 3\% at $x\approx0.2$, a smaller reduction with respect to fits with single inclusive jet data. At present we note that contrary to the case of the single-inclusive jet observable, the only available dijet measurement at $\sqrt{s}=8$ TeV is from the CMS experiment, thus making the current dijet dataset more limited than the single-inclusive dataset. 

\begin{table}[t]
\renewcommand*{\arraystretch}{1.60}
\scriptsize
\centering
\begin{tabularx}{\textwidth}{Xrccccccccc}
\toprule
 Dataset                    & $n_{\rm dat}$ &     b  &   bn   &  danw  &    d7  &   d7n  &  d7nw  &    d8  &   d8n  &  d8nw  \\
\midrule
 Jets (all)                 &        520  & [1.48] & [2.60] & [2.06] & [1.62] & [2.75] & [2.70] & [1.42] & [1.94] & [2.14] \\
 \ \ Jets (fitted)          &             &  ---   &  ---   &  ---   &  ---   &  ---   &  ---   &  ---   &  ---   &  ---   \\
 \ \ ATLAS 7 TeV            &         31  & [1.26] & [1.87] & [1.63] & [1.26] & [1.86] & [1.74] & [1.00] & [1.70] & [1.61] \\
 \ \ ATLAS 8 TeV            &        171  & [2.60] & [5.01] & [3.36] & [2.62] & [4.80] & [4.65] & [2.18] & [3.30] & [3.55] \\
 \ \ CMS   7 TeV            &        133  & [0.60] & [1.06] & [1.06] & [0.71] & [1.13] & [1.14] & [0.77] & [0.97] & [1.07] \\
 \ \ CMS   8 TeV            &        185  & [1.10] & [1.59] & [1.64] & [1.42] & [2.16] & [2.17] & [1.27] & [1.41] & [1.68] \\
 Dijets (all)               &        266  & [3.49] & [3.07] &  1.65  & [3.03] & [2.21] & [2.16] & [2.38] & [1.74] & [1.71] \\
 \ \ Dijets (fitted)        &             &  ---   &  ---   &  1.65  &  1.33  &  1.79  &  1.72  &  3.69  &  1.59  &  1.68  \\
 \ \ ATLAS 7 TeV            &         90  & [1.49] & [2.47] &  1.76  &  1.20  &  1.94  &  1.78  & [1.04] & [1.96] & [1.78] \\
 \ \ CMS   7 TeV            &         54  & [2.06] & [2.40] &  1.60  &  1.54  &  1.55  &  1.63  & [1.67] & [1.70] & [1.66] \\
 \ \ CMS   8 TeV            &        122  & [5.60] & [3.81] &  1.58  & [5.03] & [2.70] & [2.67] &  3.69  &  1.59  &  1.68  \\
\midrule
 Total                      &             &  1.20  &  1.18  &  1.22  &  1.33  &  1.20  &  1.19  &  1.33  &  1.20  &  1.20  \\
\bottomrule
\end{tabularx}

\vspace{0.3cm}
\caption{Same as Table~\ref{tab:chi2s}, but now for dijets. The baseline is repeated for ease of reference~\cite{AbdulKhalek:2020jut}.}
\label{tab:chi2d}
\end{table}

\section{Conclusions}
In this talk we presented a phenomenological study of inclusive jet production at
the LHC, exploiting recent theory calculations, in particular of NNLO QCD corrections, and compared the impact of the inclusive dijet observable,
along with the single-inclusive jet observable which is routinely 
used for PDF determination. Further details can be found in Ref.~\cite{AbdulKhalek:2020jut}.

We have found full consistency between the constraints imposed on parton distributions,
specifically the gluon, by single-inclusive jets and dijets, establishing the viability of the dijet observable PDF determination. For the case of a global NNPDF3.1-like PDF determination we observed an enhancement of the gluon distribution in the large-$x$ region
after the inclusion of either single-jet inclusive or dijet data to the baseline fit, but a stronger pull can be seen when using the dijet data. The observed pull on the gluon distribution is consistent with the CT18 analysis~\cite{AbdulKhalek:2020jut,Hou:2019efy} which includes the 8 TeV CMS single-jet inclusive data, and is consistent with fitting top data~\cite{Nocera:2017zge} which also leads to an enhancement of the gluon in the x$\geq$0.1 region.

\section*{Acknowledgments}
I would like to thank Rabah Abdul Khalek, Stefano Forte, Thomas Gehrmann, Aude Gehrmann-De Ridder, Tommaso Giani, Nigel Glover, Alexander Huss, Emanuele R. Nocera, Juan Rojo, Giovanni Stagnitto for the collaboration on the work reported here. This work was supported by the Funda\c{c}\~ao para a Ci\^encia e Tecnologia (FCT, Portugal) through the contract UIDP/50007/2020 and project CERN/FIS-PAR/0024/2019, and by the COST Action CA16201 PARTICLEFACE.

\end{document}